\documentclass[useAMS,usenatbib,twocolumn]{mn2e}
\usepackage{amsmath}
\usepackage{amssymb}
\usepackage{graphicx}
\usepackage{natbib}
\usepackage{lastpage}

\newcommand{\be}{\begin{equation}}
\newcommand{\ee}{\end{equation}}
\newcommand{\bea}{\begin{eqnarray}}
\newcommand{\eea}{\end{eqnarray}}

\newcommand{\mnras}{MNRAS}
\newcommand{\apj}{ApJ}
\newcommand{\nat}{Nature}
\newcommand{\aj}{AJ}
\newcommand{\apjl}{ApJL}
\newcommand{\aap}{A\&A}

\newcommand{\pasj}{PASJ}
\newcommand{\zp}{z/z_0}
\title{Vertical Oscillations of Fluid and Stellar Disks}
\author[L. M. Widrow and G. Bonner]{Lawrence M. Widrow$^{1}$\thanks{E-mail:\,widrow@astro.queensu.ca}
and Gage Bonner$^{2}$\footnotemark[1]\\
$^1$Department of Physics, Engineering Physics, and Astronomy, Queen's
University, Kingston, ON, K7L 3N6, Canada\\
$^2$Department of Physics, Carleton University, Ottawa, ON K1S 5B6, Canada}
\begin{document}

\date{Accepted 2015 March 12. Received 2015 March 11; in original form 2015 February 03}

\pagerange{\pageref{firstpage}--\pageref{lastpage}} \pubyear{2015}

\maketitle

\label{firstpage}

\begin{abstract}
  A satellite galaxy or dark matter subhalo that passes through a
  stellar disk may excite coherent oscillations in the disk
  perpendicular to its plane.  We determine the properties of these
  modes for various self-gravitating plane symmetric systems (Spitzer
  sheets) using the matrix method of Kalnajs.  In particular, we find
  an infinite series of modes for the case of a barotropic
  fluid.  In general, for a collisionless system, there is a double
  series of modes, which include normal modes and/or
  Landau-damped oscillations depending on the phase space distribution
  function of the stars.  Even Landau-damped oscillations may decay
  slowly enough to persist for several hundred Myr.  We discuss the
  implications of these results for the recently discovered vertical
  perturbations in the kinematics of solar neighborhood stars and for
  broader questions surrounding secular phenomena such as spiral
  structure in disk galaxies.

\end{abstract}


\section{INTRODUCTION}

In the $\Lambda$CDM cosmological paradigm, galactic disks are embedded
in extended halos of dark matter, which are populated by satellite
galaxies, star streams and dark matter subhalos.  Inevitably, some of
this halo substructure will pass through the disk, heating and
thickening the disk and also triggering the development of secular
phenomena such as spiral structure, bars, and warps.

In one of the earliest studies of disk heating, \citet{toth1992}
calculated the energy that a passing satellite deposits in a stellar
disk by assuming that the satellite gravitationally scatters
individual disk stars.  Their analysis, which was based on the
\citet{chandra1943} dynamical friction formula, suggested that
satellite infall could account for the thickness and velocity
dispersion of galactic disks.  On the other hand, the thinness and
coldness of disks could be used to set constraints on the rate of
satellite infall and hence the underlying cosmological model.
Subsequent numerical experiments confirmed that satellites can in fact
heat and thicken stellar disks, though perhaps not as efficiently as
was suggested by Toth and Ostriker (see, for example \citet{quinn1993,
  walker1996, huang1997, benson2004, gauthier2006, kazantzidis2008}).

The \citet{toth1992} calculation ignores two important effects.
First, satellites are tidally disrupted by the gravitational field of
the host galaxy, especially when their orbits take them into the
central disk-dominated region.  Second, disks can respond coherently
to the gravitational field of a satellite.  Hence, much of the energy
that is transferred to the disk takes the form of large-scale
perturbations such as warps in the outer disk \citep{quinn1993} or a
tilt of the disk plane \citep{huang1997}.  Indeed satellites can
change the morphology of a galaxy.  \citet{gauthier2006},
\citet{dubinski2008}, and \citet{kazantzidis2008} for example found
that satellites and dark matter subhalos can provoke the formation of
a bar and/or spiral structure.  More recently, \citet{purcell2011}
suggested that the Sagittarius dwarf spheroidal galaxy might be
responsible for the Milky Way's bar and spiral structure.

The importance of coherent perturbations in disk-satellite
interactions was stressed by \citet{sellwood1998}.  They argued that a
passing satellite transfers energy to the disk through the excitation
of bending waves, which eventually decay through Landau damping.  The
upshot is that since the energy transfer from satellite to disk occurs
through coherent perturbations disk heating is non-local.

\citet{toomre1966} derived the dispersion relation for bending
  waves in a stellar disk of zero thickness and uniform surface
  density.  In general, these waves propagate in the disk plane with a
  group velocity that depends on the properties of the disk and the
  wavelength of the perturbation.  Though Toomre's analysis showed
  that a perturbation with a sufficiently short wavelength is
  susceptible to a buckling instability, he argued that this
  instability would be suppressed by the random motions of the stars
  in the vertical direction.  These results were confirmed by
  \citet{araki1985} who studied wavelike perturbations of a stellar
  disk with finite thickness.  In his equilibrium model, the surface
  density as well as the velocity dispersions in the vertical and
  horizontal directions are constant in space while the vertical
  structure is given by the isothermal plane solutions of
  \citet{spitzer1942} and \citet{camm1950}.  He found that the system
  avoided the buckling instability at all wavelengths provided the
  vertical velocity dispersion was greater than $0.293$ times the
  horizontal velocity dispersion.  \citet{araki1985} also considered
  breathing modes, which correspond to compression and expansion of
  the disk perpendicular to and symmetric about the disk midplane.  In
  this case, the system can undergo a Jeans instability.

  Bending and breathing modes are the simplest modes in disklike
  systems, although they are no means only ones.  In order to explore
  vertical perturbations in more detail, one can ignore variations in
  the horizontal direction.  \citet{antonov1971,
    kalnajs1973,fridman1984}, for example, considered strictly
  vertical perturbations of a homogeneous slab and found an infinite
  double series of normal modes.  On the other hand \citet{mathur1990}
  and \citet{weinberg1991} investigated vertical oscillations in a
  variant of the Spitzer sheet where the stellar energy distribution
  is truncated (the lowered Spitzer sheet) and were able to identify
  only a handful of normal modes.

  Our aim is to carry out a comprehensive study of vertical
  oscillations in stellar and gaseous disks.  In particular we extend
  the analysis of \citet{mathur1990} and \citet{weinberg1991} to
  Landau-damped oscillations by using the matrix method of
  \citet{kalnajs1977} and complex analysis techniques from
  \citet{landau1946} and \citet{lyndenbell1962}.  We begin by
  calculating the normal modes of a gaseous disk, which serves as a
  warm-up to the the more complex collisionless case.  We then discuss
  normal modes of the homogeneous slab.  Finally we turn to the
  original (untruncated) and lowered Spitzer sheets.  In both cases,
  we identify a double series of Landau-damped oscillations.  We
  contend that these ``modes'', combined with the normal modes found
  in \citet{mathur1990} and \citet{weinberg1991}, are analogous to the
  double series of modes of the homogeneous slab.  Since we only
  consider vertical motions we cannot directly address the question of
  how modes propogate in the disk plane as was done in
  \citet{araki1985} (see, also a brief discussion in
  \citet{weinberg1991}) though we do gain a more complete
  understanding of vertical modes and Landau-damped oscillations.

The primary motivation for this work comes from recent observations of
vertical phase space structures in the kinematics of disk stars in the
solar neighborhood of the Milky Way.  These observations come free
three surveys: the Sloan Extension for Galactic Understanding and
Exploration (SEGUE; \citet{yanny2009}), the RAdial Velocity Experiment
(RAVE; \citet{steinmetz2006}), and the LAMOST Experiment for Galactic
Understanding and Exploration (LEGUE; \citet{deng2012}).  These
surveys provide full six-dimensional phase space information for tens
of thousands of stars within a few kiloparsecs of the Sun.  Recently,
several groups have detected vertical bulk motions using data from
these surveys \citep{widrow2012, williams2013, carlin2013}, which
appear to take the form of compression and expansion of the stellar
disk.  In addition, \citet{widrow2012} and \citet{yanny2013} found
evidence for wavelike North-South asymmetries in the number counts of
solar neighborhood stars.

Velocity and number density perturbations normal to the Galactic
midplane can result from satellite-disk interactions
\citep{widrow2012, gomez2013, widrow2014}.  \citet{gomez2013}, for
example, used N-body experiments to show that a Sagittarius-like dwarf
with a mass of $10^{10.5}-10^{11}\,M_\odot$ could produce density
perturbations of the same amplitude as was seen in \citet{widrow2012}
and \citet{yanny2013}.  Simulations by \citet{feldmann2015} showed
that lower mass ($10^8-10^9\,M_\odot$) satellites produce subtle
features in the bulk velocity field of the disk that might be observed
in the next generation of astrometric surveys such as Gaia
\citep{perryman2001}.

The bar and spiral structure of the Milky Way can also perturb the
velocity distribution of stars in the disk and might therefore be
responsible, at least in part, for the observations described above.
In particular \citet{dehnen2000}, \citet{fux2001} and \citet{bovy2010}
showed that a resonant interaction between the Milky Way's bar and
stars in the solar neighborhood might be responsible for the Hercules
stream, a group of co-moving stars whose bulk velocity is offset from
that of the local standard of rest.  \citet{faure2014} used
test-particle simulations to study the response of disk stars to a
spiral potential perturbation and showed that spiral structure could
generate vertical bulk motions in the disk akin to what has been
observed in the SEGUE, RAVE, and LEGUE surveys.
\citet{debattista2014} carried out a fully self-consistent N-body
simulation of a spiral galaxy and came to similar conclusions.
Essentially, a spiral arm causes compression and expansion as it
sweeps through the disk.  

Of course, since satellites have also been implicated in triggering
the formation of bars and spiral structure, it may be difficult to
disentangle perturbations that come directly from a disk-satellite
interaction and perturbations due to structures in the disk, which
themselves were the result of passing satellites.  \citet{widrow2014}
followed the evolution of bending and breathing modes in the
\citet{gauthier2006} simulation, where the disk was subjected to the
continual perturbations of a substructure-filled halo.  Bending and
breathing modes appear at early times as subhalos churn up the disk.
The formation of a bar, which occurs at about $5\,{\rm Gyr}$, is
clearly triggered by substructure-disk interactions and it may well be
that the breathing mode perturbations are the mechanism by which this
occurs.  Moreover, at late times the bar itself maintains both bending
and breathing modes, especially in the inner parts of the galaxy.

A second and more academic purpose for this work is to explore normal
modes and Landau damping in self-consistent one-dimensional systems.
In the usual textbook explanation of the Jeans instability one
considers perturbations of a spatially homogeneous mass distribution.
In the case of a fluid, one assumes that the sound speed is constant
while for a collisionless system, one assumes a Maxwellian velocity
distribution with constant velocity dispersion.  In either case, one
must confront the conundrum that the unperturbed gravitational
potential $\psi_0$ is ill-defined.  By symmetry,
$\boldsymbol{\nabla}\psi_0=0$ in a homogeneous system whereas
Poisson's equation implies $\nabla^2\psi_0 = 4\pi G\rho_0$.  These two
equations are inconsistent.  The Jeans swindle, wherein one makes the
{\it ad hoc} assumption that Poisson's equation applies only to
perturbed quantities, provides a way forward (see \citet{binney2008} for
a more detailed discussion).  The linearized equations are then
solved by making the ansatz that the perturbed density and potential
vary harmonically in space and time.  In both the fluid and
collisionless cases a perturbation whose wavelength is longer than the
Jeans length grows exponentially.  A perturbation in a fluid whose
wavelength is less than the Jeans length oscillates as sound waves.
On the other hand, a short wavelength perturbation in a collisionless
system undergoes Landau damping and rapidly decays.  In the models
considered here, the density is concentrated in the midplane and the
Poisson equation can be solved without resorting to the Jeans swindle.

The outline of the paper is as follows: In Section 2 we find the
normal modes for an isothermal plane-symmetric fluid.  In Section 3,
we derive normal and Landau-damped oscillations for the three examples of
collisionless systems mentioned above.  In Section 4, we present
results from a simulations of a one-dimensional system that exhibits
damping in a perturbed Spitzer sheet.  We conclude in Section 5 with a
summary and discussion of our results and some thoughts on further
directions for this line of research.  Three appendices provide
mathematical details for some of our calculations.

\section{LINEAR PERTURBATIONS OF AN ISOTHERMAL ONE-DIMENSIONAL FLUID}

In this section and the ones that follow, we consider the vertical
perturbations of plane symmetric systems.  The idea of treating a
galactic disk as a plane symmetric system can be traced to the seminal
paper by \citet{oort1932}, wherein the motions of stars perpendicular
to the Galactic midplane were used to estimate the vertical force and
mass distribution in the solar neighborhood.  \citet{spitzer1942}
derived an equilibrium model for a self-gravitating, plane-symmetric
system of stars under the assumption that the stellar velocity
dispersion is constant with height above the midplane.  His derivation
is based on the Jeans equations (i.e., moments of the collisionless
Boltzmann equation) and is therefore akin to the derivation of the
equilibrium fluid model considered in this section.  \citet{camm1950}
solved for the equilibrium distribution function, which provides the
starting point for our analysis in Section 3.  In what follows, we
refer to these models collectively as Spitzer sheets.

We consider linear perturbations in a simple self-gravitating
barotropic fluid.  For an analysis of perturbations in a multiphase,
magnetized model of the interstellar medium see \citet{walters2001}.
A fluid with density $\rho$, velocity $v$, pressure $p$, and
gravitational potential $\psi$ obeys the continuity, Euler, and
Poisson equations.  For a plane symmetric system, these equations
become

\begin{equation}
\frac{\partial \rho}{\partial t} + \frac{\partial}{\partial z}\left (\rho v\right ) = 0
\end{equation}

\begin{equation}
\frac{\partial v}{\partial t} + v\frac{\partial v}{\partial z} 
= -\frac{1}{\rho}\frac{\partial p}{\partial z} - \frac{\partial \psi}{\partial z}
\end{equation}

\begin{equation}
\frac{\partial^2\psi}{\partial z^2} = 4\pi G \rho
\end{equation}

\noindent where we choose our coordinate system so that the $z$-axis
is normal to the symmetry plane.  We assume that the fluid has an
equation of state $p=p\left (\rho\right )$ and constant sound speed
$v_s \equiv \left (dp/d\rho\right )^{1/2}$.

We write $\rho$, $v$, $p$, and $\psi$ as the sum of an
equilibrium solution and a linear perturbation, e.g.,

\begin{equation}
\rho = \rho_0\left (z\right ) + \rho_1\left (z,t\right )~.
\end{equation}

\noindent For the equilibrium solution, $v_0 = 0$ and the continuity
equation is satisfied automatically while the Euler and Poisson
equations are solved by the following density-potential pair:

\be\label{eq:psi0}
\psi_0\left (z\right ) = 2v_s^2\ln\left (\cosh\left (\zp\right )\right )
\ee

\noindent and

\be\label{eq:rho0}
\rho_0\left (z\right ) = \rho_c\,{\rm sech}^2\left (\zp\right )
\ee

\noindent where $z_0 \equiv v_s/\left (2\pi G\rho_c\right )^{1/2}$ and
$\rho_c$ is the density in the midplane \citep{spitzer1942, camm1950}.

In what follows, we use a system of units in which $z_0 = v_s = G = 1$
and $\rho_c = 1/2\pi$.  The linearized equations are

\begin{equation}
\frac{\partial \rho_1}{\partial t} + \frac{\partial}{\partial z}\left (\rho_0 v_1\right ) = 0
\end{equation}

\begin{equation}
\rho_0\frac{\partial v_1}{\partial t} = -\frac{\partial \rho_1}{\partial z} 
- \rho_1\frac{\partial \psi_0}{\partial z}
- \rho_0\frac{\partial \psi_1}{\partial z}
\end{equation}

\begin{equation}
\frac{\partial^2\psi_1}{\partial z^2} = 4\pi \rho_1~.
\end{equation}

\noindent To search for modes, we assume that the first-order
quantities are proportional to $\exp{(-i\omega t)}$ (e.g.,
$\rho_1\left (z,\,t\right ) = \exp\left ({-i\omega t}\right
)\tilde{\rho_1}(z)$) and combine the continuity and Euler equations to
arrive at a single equation for $\tilde{\rho_1}$:

\begin{align}
\label{eq:Eulerpert}
\omega^2\tilde{\rho_1}  = 
-\frac{d^2\tilde{\rho_1}}{dz^2} 
& - \frac{d\tilde{\rho_1}}{dz}\frac{d\psi_0}{dz} \\ 
& - \tilde{\rho_1}\frac{d^2\psi_0}{dz^2}
- \frac{d\rho_0}{dz}
\frac{d\tilde{\psi_1}}{dz}
- \rho_0\frac{d^2\tilde{\psi_1}}{dz^2}~.
\end{align}
 
\noindent For a homogeneous system, the fourth term on the right-hand
side is zero while the second and third terms are ignored by
implementing the Jeans swindle \citep{binney2008}.  We then assume a
sinusoidal spatial dependence for the perturbation
($\tilde{\rho}_1,~\tilde{\psi}_1\propto \exp{(ikz)}$) and use the
first-order Poisson equation to arrive at the dispersion relation

\begin{equation}
\omega^2 = v_s^2\left (k^2 - k_J^2\right) = k^2 - 2
\end{equation}

\noindent where $k_J \equiv \left (4\pi G\rho_c/v_s^2\right )^{1/2} =
2^{1/2}$ is the Jeans wavenumber.  For $k<k_J$ (long wavelength
perturbations) $\omega$ is imaginary signaling an
instability with growth rate $\alpha = \left (2-k^2\right )^{1/2}$.
On the other hand, for $k>k_J$ perturbations oscillate with
frequency $\omega = \left (k^2-2\right )^{1/2}$.

To study linear perturbations of the Spitzer sheet we use the Kalnajs matrix
method and  write $\tilde{\rho}_1$ and $\tilde{\psi}_1$ in terms of a biorthonormal 
basis \citep{kalnajs1971, kalnajs1977, binney2008}:

\begin{equation}\label{eq:expansion}
\tilde{\psi}_1\left (z\right ) = \sum_j c_j \psi_j\left (z\right )
~~~~~~
\tilde{\rho}_1\left (z\right ) = \sum_j c_j \rho_j\left (z\right )
\end{equation}
\noindent where
\begin{equation}\label{eq:poissonexpansion}
\frac{d^2\psi_j}{dz^2} = 4\pi \rho_j
\end{equation}
and
\begin{equation}\label{eq:normalization}
\int_{-\infty}^\infty dz\,\psi_j^*(z)\rho_k(z) = -\delta_{jk}~.
\end{equation}

\noindent We multiply Eq.\,\ref{eq:Eulerpert} by $-\psi_j$ and
integrate with respect to $z$ from $-\infty$ to $\infty$ to obtain a
matrix equation of the form

\begin{equation}\label{eq:matrix}
\omega^2 c_j = \sum_k M_{jk} c_k~.
\end{equation}

\noindent Evidently, the eigenvalues of ${\bf M}$ are the 
squares of the mode frequencies.

For the calculation at hand we use the basis introduced in \citet{araki1985}:
\begin{equation}
\psi_j\left (z\right ) = N_jP_j(u)
\end{equation}
\noindent and
\begin{equation}
\rho_j\left (z\right ) = 
-N_j\frac{j\left (j+1\right )}{4\pi}
\left (1-u^2\right )P_j(u)
\end{equation}
\noindent where $u = {\rm tanh}(z)$, $P_j$ are Legendre polynomials,
and $N_j\equiv \left (2\pi\left (2j+1\right )/{j\left (j+1\right )}\right
)^{1/2}$ are normalization constants.  The matrix elements $M_{jk}$
can be calculated analytically (see Appendix A).  Note that $M_{jk}$
is nonzero if and only if $j=k-2,\,k,$ or $k+2$.  The matrix therefore
separates into two independent matrices, one where $j$ and $k$ are
both even and the other where $j$ and $k$ are both odd.  The physical
implication is that modes have definite parity, a property that
follows from the symmetry of the equilibrium system under the
transformation $z\to -z$.

\begin{figure}
\centering
\includegraphics[width=.5\textwidth]{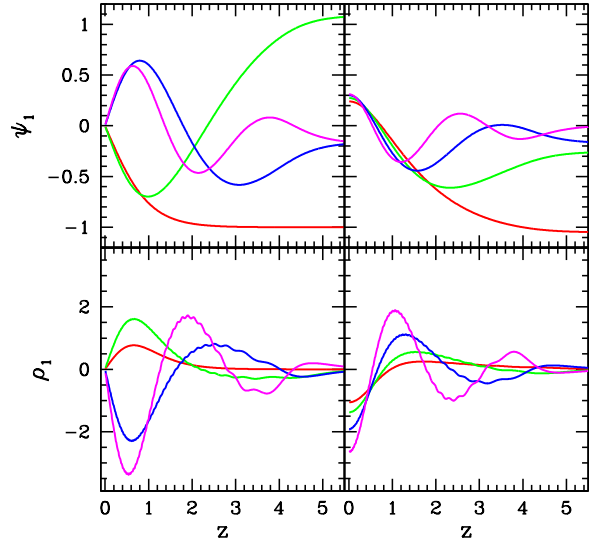}
\caption{Normal modes of the fluid Spitzer sheet.
Panels on the left show the potential (top) and density (bottom) for the
odd parity modes with $n=1$ (red curve), $n=3$ (green), $n=5$ (blue)
and $n=7$ (magenta).  Panels on the right show the even
parity modes $n=2$ (red curve), $n=4$ (green), $n=6$ (blue)
and $n=8$ (magenta).}
\label{fig:fluidmodes}
\end{figure}

The eigenvalues and eigenvectors for these sparse matrices are found
using the software package LAPACK \citep{anderson1992}.  The
frequencies $\omega_n$ and eigenfunctions for the lowest eight modes
are given in Table 1 and Figure \ref{fig:fluidmodes}.  The
eigenfunction label $n$ corresponds to the number of nodes in the density
profile.  Note that the $n=1$ mode, which has zero frequency, corresponds
to a shift in the system as a whole.

\begin{table*}
\begin{tabular}{cccc}
\hline
odd modes & &  even modes\\
$n$ & $\omega_{n}$ & $n$ & $\omega_{n}$\\
1 & 0& 2 & 1.129\\
3 & 1.522 & 4 & 2.180  \\
5 & 3.138 & 6 & 4.373\\
7 & 5.976 & 8 & 7.865\\
\hline
\end{tabular}
\caption{Frequencies, in units of where $v_s=z_0=1$ for the first four even
and first four odd modes.}
\end{table*}

\section{COLLISIONLESS SYSTEMS}

\subsection{Formalism}

The dynamics of a collisionless, plane symmetric system is described
by the collisionless Boltzmann and Poisson equations in one dimension.
We follow the formalism found in \citet{mathur1990} and
\citet{weinberg1991} in which the collisionless Boltzmann equation is
written in terms of angle-action variables (see, also
\citet{binney2008}).  For an alternate approach based on Jeans
equations, see \citet{louis1992}.  A particle in a time-independent
potential, $\psi_0(z)$ executes periodic motion with constant energy
$E = v^2/2 + \psi_0\left (z\right )$, period $T(E)$, and maximum
excursion from the midplane $z_{\rm max}$ where $\psi_0\left (z_{\rm
    max}\right ) = E$.  We can therefore introduce the angle-action
variables $\left (\theta,\,E\right )$ where $\theta$ is defined so
that $dt = T(E) d\theta/2\pi$ and $\theta = 0$ corresponds to
$\left (z,\,v\right ) =\left (-z_{\rm max},\, 0\right )$.

As before, we write the density and potential, along with the
distribution function, as the sum of an equilibrium solution and a
linear perturbation.  For example
\begin{equation}
f\left (E,\,\theta,\,t\right ) = f_0\left (E\right ) + f_1\left
  (E,\,\theta,\,t\right )~.
\end{equation}

\noindent The linearized collisionless Boltzmann and Poisson
equations are then

\begin{equation}\label{eq:linearCBE}
\frac{\partial f_1}{\partial t} + \frac{2\pi}{T(E)}
\frac{\partial f_1}{\partial \theta} - 
\frac{2\pi}{T(E)}
\frac{\partial f_0}{\partial E}
\frac{\partial \psi_1}{\partial \theta} = 0
\end{equation}

\noindent and

\begin{equation}\label{eq:linearPoisson}
\frac{\partial^2 \psi_1}{\partial z^2} = 4\pi G\int dv\,f_1~.
\end{equation}

\noindent Note that the Newtonian potential that appears in the
collisionless Boltzmann equation includes contributions from both the
system and any external perturbations while $f_1$ refers only to the
system.

Since particle orbits are periodic in $\theta$ we can expand each of
the first-order quantities in a Fourier series, e.g., 

\begin{equation}
f_1\left (E,\,\theta, t\right ) = \sum_{n=-\infty}^\infty
f_n\left (E,\,t\right
)e^{in\theta}~.
\end{equation}
Eq. \ref{eq:linearCBE} is satisfied for each Fourier component:

\begin{equation}\label{eq:fouriercbe}
\frac{\partial f_n}{\partial t} + \frac{2\pi in}{T(E)}f_n
-\frac{2\pi i n}{T(E)}
\frac{\partial f_0}{\partial E}\psi_n  = 0~.
\end{equation}
\noindent The temporal Fourier transform of $f_n$ is 
\begin{equation}
\tilde{f}_n\left (E,\,\omega\right ) = 
\int_0^\infty dt\,f_n\left (E,\,t\right ) e^{i\omega t}
\end{equation}
and similarly for $\psi_n$.  In writing $\tilde{f}_n$, we have
  assumed that $f_n=0$ for $t<0$ and that there exists a real number
  $\gamma>0$ such that $\int dt\,{\rm exp}\left (-\gamma t\right )
  f_n(t)$ converges.  The latter condition insures the existence of
  the inverse Fourier transform (\citet{binney2008} -- see
  Eq.\,\ref{inverseFT} below.  We multiply Eq.\,\ref{eq:fouriercbe}
by ${\rm exp}\left (i\omega t\right )$ and integrate over $t$.  The
first term is handled by an integration by parts and we obtain an
algebraic equation for $\tilde{f}_n$, which can be written as

\begin{equation}\label{eq:tildefn}
\tilde{f}_n\left (E,\,\omega\right ) = \frac{d f_0}{d E}
\frac{n\Omega(E)\tilde{\psi}_n\left (E,\,\omega\right )}{n\Omega(E) -
  \omega}
\end{equation}

\noindent where $\Omega(E) \equiv 2\pi/T(E)$ is the orbital frequency
for a particle with energy $E$.

As in the fluid case, we decompose the density and potential in terms
of a biorthonormal basis.  Let $\psi_1^s$ be the linear perturbation
to the potential due to the system itself and $\psi_1^e$ be the
linear external potential (if one exists).  For the system, we have

\begin{equation}
\tilde{\psi}_1^s\left (z,\,\omega\right )  = 
\sum_{j=1}^{\infty} c_j\left (\omega\right )\psi_j\left (z\right )
\end{equation}

\noindent and

\begin{equation}
\tilde{\rho}_1^s = \int dv\,f_1 = \sum_{j=1}^\infty c_j\left (\omega\right )
\rho_j\left (z\right )
\end{equation}

\noindent where the potential-density pair $\left (\psi_j,\,\rho_j\right )$
satisfy Eqs.\,\ref{eq:expansion}-\ref{eq:normalization}.  Likewise, 
for the external potential, we have

\begin{equation}
\tilde{\psi}_1^e\left (z,\,\omega\right )   
 = \sum_{j=1}^\infty d_j\left (\omega\right )\psi_j\left (z\right )~.
\end{equation}

\noindent We then write $\psi_j$ as a Fourier series

\begin{equation}
\psi_j(z) = \sum_{n=-\infty}^\infty \psi_{jn}(E)e^{in\theta}
\end{equation}

\noindent where
\begin{equation}\label{eq:psibasisFT}
\psi_{nj}(E) = 
\frac{1}{2\pi}
\int_0^{2\pi} d\theta \,
\psi_j\left (z\left (E,\,\theta\right ) \right ) e^{-in\theta}~.
\end{equation}

\noindent We now use Eq.\,\ref{eq:tildefn} and the expansions defined above
to write the Poisson equation as follows:

\begin{align}
\frac{1}{4\pi G} 
\frac{d^2\tilde{\psi}_1}{dz^2} 
& = \sum_{j=1}^\infty c_j\rho_j \nonumber\\
& = \sum_{n=-\infty}^\infty \int dv \frac{df_0}{dE}
\frac{n\Omega \tilde{\psi}_n}{n\Omega - \omega} e^{in\theta}
\end{align}

\noindent where $\tilde{\psi}_n = 
\tilde{\psi}_n^s +  
\tilde{\psi}_n^e = \sum_j\left (c_j + d_j\right )\psi_{jn}$. 

We multiply this equation by $-\psi_j(z)$ and 
integrate over $z$ to arrive at a matrix equation the expansion
coefficients $c_j$:

\begin{equation}\label{eq:matrixequation}
c_j\left (\omega\right ) 
= M_{jk} \left (\omega\right )
\left (c_k\left (\omega\right ) + d_k \left (\omega\right )\right )
\end{equation} 
where
\begin{equation}\label{eq:matrixelements}
M_{jk}\left (\omega\right ) = -4\pi\sum_{n=1}^{\infty}
\int dE
\frac{df_0}{dE}\frac{\Omega\psi_{jn}\psi_{k n}}
{\Omega^2 - \omega^2/n^2}
\end{equation}

\noindent \citep{mathur1990,weinberg1991}.  Note that in deriving
Eq.\,\ref{eq:matrixelements}, we have combined positive and negative
values of $n$ and also omitted the $n=0$ term as its contribution to
the sum is zero.  The matrix ${\bf M}$, which is sometimes referred to
as the polarization matrix \citep{binney2008}, describes the response
of the system to the total perturbing potential (system plus
external).  We can rewrite Eq.\,\ref{eq:matrixequation} as an explicit
equation for the $c_j$:

\begin{equation}  
c_j\left (\omega\right ) = \sum_k
R_{jk}\left (\omega\right )  
d_k\left (\omega\right )
\end{equation} 
where ${\bf R} \equiv \left ({\bf I-M}\right )^{-1}{\bf M}$
describes the response of the system to an external potential.

The time-dependent response of the system is determined by
calculating the inverse Fourier transform of $c_j$:

\begin{align}\label{inverseFT}
c_j\left (t\right )  & = \frac{1}{2\pi}
\int_{i\gamma-\infty}^{i\gamma + \infty} d\omega \,c_j\left (\omega\right
)e^{-i\omega t}  \\
& = -i\sum_r c_{j,r}\left (\omega_r\right ) e^{-i\omega_r t}
\end{align}
where $\omega_r$ are the poles of the function $c_j\left (\omega\right
)$.  These poles occur at the poles of the response matrix ${\bf R}$
or equivalently, where either ${\rm det}({\bf M})$ is singular or
${\rm det}\left ({\bf I}-{\bf M}\right )=0$.  The second of these
conditions is equivalent to the eigenvalue equation
\begin{equation}\label{eq:selfconsistent} 
c_j\left (\omega\right ) = \sum_k M_{jk}\left (\omega\right ) 
c_k\left (\omega\right )~.
\end{equation}
Our goal is therefore to find frequencies $\omega_m$ such that one of
the eigenvalues of ${\bf M}$ is equal to unity.  The corresponding
functions $\psi_1^s\left (z,\,\omega_m\right )$ and $\rho_1^s\left
(z,\,\omega_m\right )$ are then the potential-density pairs for either
normal modes, when ${\rm Im}\left (\omega_m\right )=0$ or
Landau-damped oscillations, when ${\rm Im}\left (\omega_m\right )<0$.
 
\subsection{Homogeneous slab of stars}

We first determine the normal modes of the spatially homogeneous slab
\citep{antonov1971, kalnajs1973, fridman1984}, where the distribution
function for the equilibrium system is given by

\begin{equation}\label{eq:kalnajsDF}
f_0(E) = 
\begin{cases}
f_c\left (1-E/\sigma^2\right )^{-1/2} &
{\rm for} ~~0 < E < \sigma^2 \\
0 & {\rm otherwise}~.
\end{cases}
\end{equation}

\noindent The density is equal to the constant $\rho_c =
2^{1/2}\pi\sigma f_c$ for $|z|<\left (\sigma^2/2\pi G\rho_c\right
)^{1/2}$ and zero otherwise.  All particles execute simple harmonic
motion about the midplane with frequency $\Omega_c = \left (4\pi
  G\rho_c\right )^{1/2}$.  We choose units akin to those used in the 
previous section, namely $\sigma = G = 1$ and $\rho_c = 1/2\pi$.
The edge of the slab is then at $z = \pm 1$ and $\Omega_c = 2^{1/2}$.  

Following \citet{kalnajs1973} we use the biorthonormal basis to describe
the density and potential inside the slab:

\be 
\psi_j(z) = N_j\left (P_{j+1}(z) - P_{j-1}(z)\right ) 
\ee 

\noindent and 

\be
\rho_j(z) = -\frac{N_j}{4\pi} \frac{j\left (j+1\right )}{1-z^2}\left
  (P_{j+1}(z) - P_{j-1}\right ) 
\ee 

\noindent where $N_j = \left (2\pi/\left (2j+1\right )\right )^{1/2}$.
In fact, $\left (\psi_j,\,\rho_j\right )$ {\it are} the normal modes
of the system.  We see that the density and potential have even parity
when $j$ is odd and vice versa.  Furthermore, when $j$ is odd, there
are $\left (j+1\right )/2$ distinct modes \citep{kalnajs1973}.  We
label the frequencies for these modes $\omega_{j,k}$ where $k =
1,\,3\dots \left (j+1\right )/2$.  Likewise, when $j$ is even, there
are $\left (j+2\right )/2$ modes.  For example, there is a single mode
with $j=1$: $\psi_1 = \left (3N_1/2\right )\left (z^2-1\right )$,
$\rho_1 = 3N_1/4\pi$, and $\omega_{1,1} = 6^{1/2} = 1.73\Omega_c$
\citep{antonov1971, kalnajs1973, fridman1984}.  The distribution
function for this mode is

\be f_1\left (z,\,v\right ) = \frac{3N_1
  E}{2} \,\frac{df_0}{dE}\,\frac{\cos{2\theta}}{2-\omega^2/4} 
\ee

\begin{figure}
\centering
\includegraphics[width=0.5\textwidth,height=8cm]{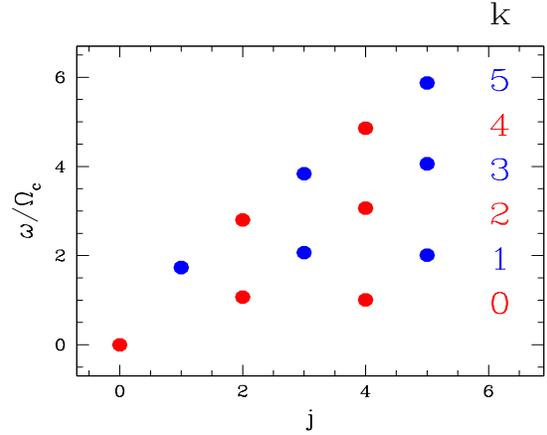}
\caption{Frequencies for normal modes of the homogeneous stellar slab.
  The dots give the frequencies from \citet{kalnajs1973} for different
  values of the eigenvalues $j$ and $k$.  Red dots are for even $j$
  and correspond to modes where the density and potential have odd
  parity under $z\to -z$.  Blue does are for even parity (odd $j$)
  modes.  Modes in the same column have the same density and potential
  but different phase space distribution functions.}
\label{fig:kalnajsfrequencies}
\end{figure}

\noindent (see Appendix B).  The frequencies of the $j\le 5$ modes are
depicted in Figure \ref{fig:kalnajsfrequencies}; numerical values can be
found in \citet{kalnajs1973}.

Self-consistency requires $\rho_1 = \int dv\,f_1$.  Formally, this
integral is infinite because of the strong divergence of $df_0/dE$ as
$E\to 1$.  To handle this divergence we consider a family of functions
$f_\lambda(E)$ that are continuous, that vanish as $E\to \infty$ and
that approach Eq.\,\ref{eq:kalnajsDF} in the limit $\lambda\to
\infty$.  We can then perform an integration by parts and afterwards
let $\lambda\to \infty$ (see for example, \citet{fridman1984}).  As
shown in Appendix B, this procedure leads to the required relation
between $\rho_1$ and $f_1$.  A similar trick can be used to evaluate
the matrix elements in Eq.\,\ref{eq:matrixelements} where we can write

\begin{equation}\label{eq:kalnajsmatrixelements}
M_{jk}\left (\omega\right ) = -
\sum_n
\frac{4\pi\Omega_c}
{\Omega_c^2 - \omega^2/n^2}\,
\int dE\,f_0
\frac{d}{dE}\left (\psi_{j n}\psi_{k n}\right ).
\end{equation}

\noindent The function $f_0$ has an integrable singularity at
$E=1$ and the matrix elements can be calculated numerically.
In doing so, we find that ${\bf M}$ is indeed diagonal, a result that
\citet{kalnajs1973} showed using various relations between Legendre
polynomials and hypergeometric functions.

\begin{figure}
\centering
\includegraphics[width=0.5\textwidth]{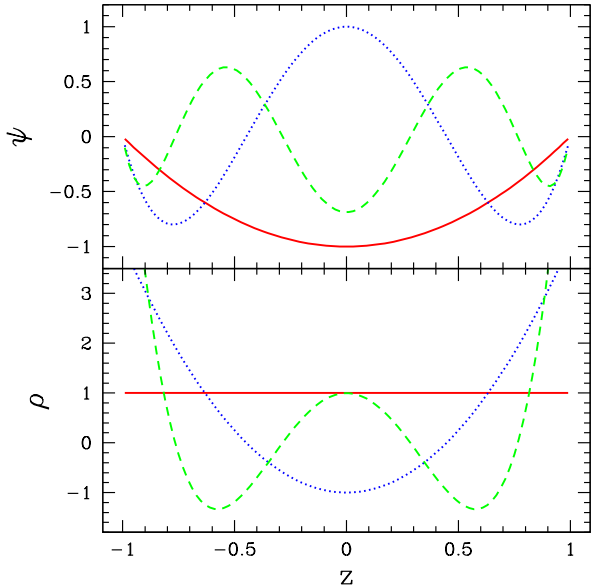}
\caption{Even parity normal modes of the homogeneous slab.
The linear potential (top panel) and density (bottom panel) are shown as a
  function of $z$ for $j=1$ (red/solid), $j=3$ (blue/dotted) and $j=5$ (green/dashed).}
\label{fig:slabpsirho}
\end{figure}

\begin{figure}
\centering
\includegraphics[width=0.5\textwidth]{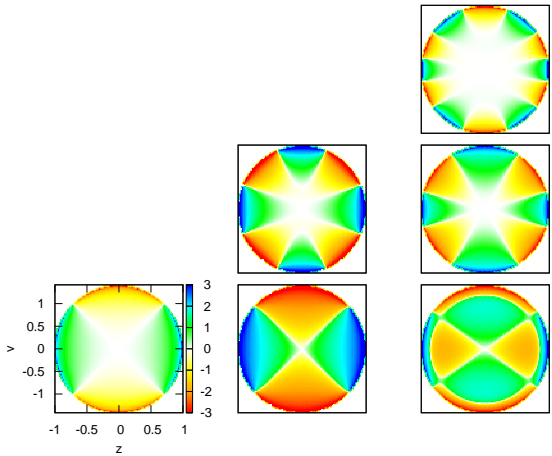}
\caption{Distribution functions for even parity modes of the
  homogeneous slab with $j=1,\,3,\,5$.  The arrangement of the panels
  is the same as the arrangement of the blue dots in Figure
  \ref{fig:kalnajsfrequencies}.  Thus, the three modes along the
  bottom row have, from left to right, $j=1,\,3,$ and $5$ and $\omega
  \simeq 1.73,\,2.07,$ and $2.01$.  The three modes along the right
  column all have $j=5$ and have a density profile and potential given
  by the green/dashed curves in Figure \ref{fig:slabpsirho}.}
\label{fig:slab}
\end{figure}

In Figure \ref{fig:slabpsirho} we show the density and potential pairs
for the even parity modes with $j=1,\,3,\,5$.  The case $j=1$ is a
breathing mode in which the density within the slab depends on $t$
but not $z$, while the boundaries of the slab move in and out
accordingly.  The distribution functions for the six distinct modes
with these values of $j$ are shown in Figure \ref{fig:slab}.  Modes in
the same row have frequencies that cluster about $\left (j+1\right
)\Omega_c$, as in Figure \ref{fig:kalnajsfrequencies}.  For example,
the three modes along the bottom row have frequencies from left to
right of $\omega/\Omega_c = 1.73, 2.07, 2.01$.  Particles make roughly
half an orbit in the time it takes the system to execute a single mode
oscillation of this type.  The frequencies for the two modes in the
middle row are $\omega/\Omega_c = 3.83,\,4.06$ so that particles make
roughly one quarter of an orbit during one oscillation period.  Modes
in the same column have the same density profile and potential since
they have the same eigenvalue $j$.

\subsection{Linear perturbations of the stellar Spitzer sheet}

In a collisionless system, oscillations that are in resonance
with any of the particles in the distribution will decay via Landau
damping whereby coherent energy in the oscillations is transferred to
energy in random particle motion.  The homogeneous slab is a special
example where all particles have the same orbital frequency
$\Omega_c$.  Since the mode frequencies are all slightly off resonance
there is no Landau damping, at least not formally.

\begin{figure}
\centering
\includegraphics[width=0.5\textwidth]{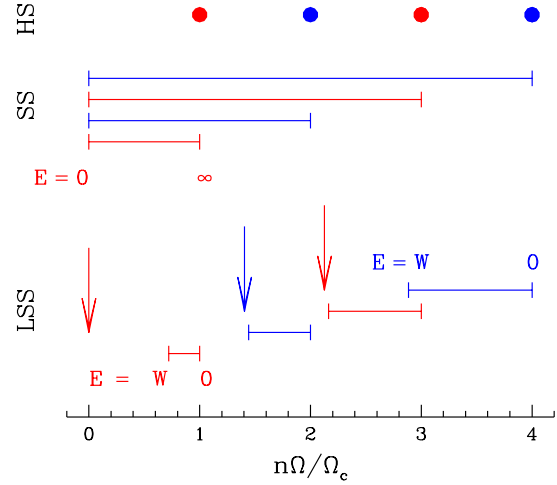}
\caption{Range of allowed frequencies for orbits of stars and their
  harmonics in the lowered Spitzer sheet (LSS), the Spitzer sheet of
  \citet{spitzer1942} and \citet{camm1950} (SS) and the homogeneous
  plane (HS).  The left-most horizontal bar in the row marked LSS
  shows the range of allowed frequencies for orbits in the lowered
  Spitzer sheet with $W=2$.  The blue bar immediately to the right of
  this shows the frequencies for the second harmonics of these orbits,
  and so on.  The vertical arrows mark the positions of the modes
  found by \citet{mathur1990} and \citet{weinberg1991}.  The frequency
  ranges for the Spitzer sheet are displayed in the same way while for
  the homogeneous plane all particles orbit at the same frequency.
  Odd parity modes and frequencies are shown in red while even parity
  modes and frequencies are shown in blue.}
\label{fig:gaps}
\end{figure}

In this section we consider the stellar Spitzer sheet where the
distribution function is
\begin{equation}\label{eq:spitzerdf}
f_0(E) = f_ce^{-E/\sigma^2}~.
\end{equation}  
The density and potential have the same form as in the fluid case
(Eqs.\,\ref{eq:psi0} and \ref{eq:rho0}) but with $v_s$ replaced by
$\sigma$ and $f_c = \rho_c/\left (2\pi\sigma^2\right )^{1/2}$.  
As before, we choose units in which $G=z_0 = \sigma = 1$.  In these
units, the orbital frequencies range from $\Omega(E)\to \Omega_c =
2^{1/2}$ for $E\to 0$ to $\Omega(E) \to 0$ for $E\to \infty$.  Thus,
coherent oscillations with frequency $\omega$ will be in resonance
with particles whose orbital frequencies are $\Omega_n = \omega/n$
where $n = n_{\rm min},\,n_{\rm min}+1,\,\dots$ and $n_{\rm min} =
{\rm Int}(\Omega_c/\omega)$.  This point is illustrated in Figure
\ref{fig:gaps}.  The line segments in the middle portion of the
diagram show the range in orbital frequencies and higher harmonics for
the Spitzer sheet.  The dots along the top portion of the diagram show
the same for the homogeneous slab.  The modes of the homogeneous slab
are all off-resonance, that is, reside in the gaps between the dots.
No such gaps exist in the case of the Spitzer sheet.  The implication
is that all coherent oscillations of the Spitzer sheet are damped.

We also consider the lowered Spitzer sheet, which provides a link
between the the homogeneous slab and the Spitzer sheet.  This model
the one-dimensional analog of the lowered isothermal sphere or King
model and has a distribution function that is given by
\begin{equation} 
f_0(E) = 
\begin{cases}
f_W \left (e^{-E/\sigma^2} - e^{-W/\sigma^2}\right ) & 0 < E < W \\ 
0 & {\rm otherwise}~.
\end{cases}
\end{equation} 

\noindent With $\sigma = 1$ and

\begin{equation}
f_W = \sigma^{-1}
\left (\left (2\pi\right )^{3/2}
{\rm erf}\left (y_0\right ) - 2^{5/2}\pi y_0e^{-y_0^2}
\right )^{-1}
\end{equation}

\noindent where $y_0\equiv W^{1/2}/\sigma$, the density in the
midplane is $\rho_c = 1/2\pi$ and the maximum orbital frequency is
$\Omega_c=2^{1/2}$ (See Appendix C2 for details).  For finite $W$
there is a lower bound on the particle frequencies: $\Omega_{\rm min}
\equiv \Omega(W)$.

The horizontal line segments in the lower portion of Figure
\ref{fig:gaps} show the range in frequencies and higher harmonics for
the case $W=2$.  We see that there are no resonant particles for $0 <
\omega \la 0.72\,\Omega_c$.  \citet{mathur1990} and
\citet{weinberg1991} refer to this range in frequency as the principal
gap.  Likewise, there are no particles with $\Omega_c<\omega \la
1.45\,\Omega_c$ (the first gap).

As discussed above, our aim is to find values of $\omega$ for which
one of the eigenvalues of the matrix ${\bf M}$ is equal to one.
\citet{mathur1990} and \citet{weinberg1991} restrict their search for
modes to real values $\omega$ that reside in the frequency gaps.  In
doing so, they avoid any potential singularities in the energy
integrals necessary for calculating the matrix elements.  The vertical
arrows in Figure \ref{fig:gaps} indicate the positions of the modes
found by the method outlined in \citet{mathur1990} and
\citet{weinberg1991}.  Note that apart from the $\omega=0$ mode, the
frequencies are slightly less than $n\Omega_{\rm min}$.

Next we consider complex values of $\omega$.  Recall that the Fourier
transform $\tilde{f}_n$ and hence the matrix elements $M_{jk}$ are
defined for ${\rm Im}(\omega) > 0$.  In this regime, the energy
integrals in Eq.\,\ref{eq:matrixelements} can be calculated by a
straightforward numerical integration since the singularity in the
integrand occurs in the lower half of the complex $E$ plane and is
avoided as one integrates along the real $E$ axis.

To explore solutions with ${\rm Im}(\omega) \le 0$ we analytically
continue ${\bf M}$ into this region of the complex $\omega$ plane.  To
see how this works we define ${\cal M}_{jk}\left
  (\omega_R,\,\omega_I;\lambda\right ) \equiv M_{jk}\left (\omega_R +
  i\lambda \omega_I\right )$ to be a function of the real variable
$\lambda$ with $\omega_R$ and $\omega_I$ taken to be positive
constants.  The analytic continuation of $M_{jk}$ will have the
property that ${\cal M}_{jk}$ is a continuous function of $\lambda$.
When $\lambda$ changes from positive to negative values, the
singularity in the energy integral crosses from the lower half of the
complex energy plane to the upper half.  Thus, if $M_{jk}$ is to be
continous when this happens, the integration contour for the energy
integral must deformed so that the it always remains above the
singularity.  This deformed contour is analogous to the so-called
Landau contour used to carry out the velocity-space integral in the
usual derivation of Landau damping for a homogeneous system
\citep{landau1946, lyndenbell1962, binney2008}.  We then have two
contributions to the energy integrals in Eq.\,\ref{eq:matrixelements},
one from the principal part of the integral, and the other from the
residue due to the singularity at $\Omega(E) = \omega/n$.
That is, each term of the sum in Eq.\,\ref{eq:matrixelements}
becomes

\begin{equation}
-4\pi{\cal P} \int dE\,
\frac{\Omega\psi_{jn}\psi_{kn}}{\Omega^2 - \omega^2/n^2}
\frac{df_0}{dE} 
+ 4\pi^2 i\left (\psi_{jn}\psi_{kn} \frac{dE}{d\Omega}
  \frac{df_0}{dE}\right )_{\Omega = \omega/n}
\end{equation}

\noindent The second term requires us to treat the quantities inside
the parenthesis as complex analytic functions of complex energy.
Details of how this is done can be found in Appendix C.

\begin{figure}
\centering
\includegraphics[width=0.65\textwidth]{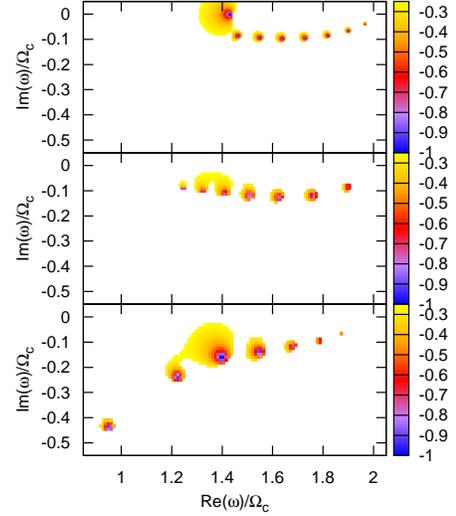}
\caption{Heat map showing the position of (Landau) modes in the
 complex frequency plane for the lowered Spitzer sheet with
 $W=2$ (top panel) and $W=5$ (middle panel) and for the isothermal
 plane (bottom panel).  Colors indicate log-base 10 of the eigenvalue
 of the matrix $M_{ij}-\delta_{ij}$ that comes closest to zero.}
\label{fig:landau}
\end{figure}

We evaluate ${\bf M}$ as a function of complex $\omega$ for $0 < {\rm
  Re}(\omega) < 8$ and $-0.6 < {\rm Im}(\omega) < 0$.  For each
$\omega$ we determine the eigenvalues of ${\bf M}$ using LAPACK
\citep{anderson1992} and find the eigenvalue closest to unity,
$\lambda^*$ Figure \ref{fig:landau} shows heat maps of $H \equiv{\rm
  log}(|\lambda^* -1|)$ as a function of $\omega$ for a lowered
Spitzer sheet with $W=2$ and $W=5$ and for the infinity (i.e., $W\to
\infty$) Spitzer sheet.  Places where $H$ is a large negative number
indicate mode frequencies.  The matrix ${\bf M}$ is truncated at
$j=8$, which explains why only $7$ or $8$ candidate modes are found.
In the $W=2$ case, there is a single undamped breathing mode with
$\omega \simeq 1.4\Omega_c$ as well as seven damped oscillations.  For the
latter, the ratio of the exponential decay time to the
oscillation period is $\tau/T = {\rm Re}(\omega)/2\pi {\rm Im}(\omega)
\simeq 2.5-8$.

\begin{figure}
\centering
\includegraphics[width=0.5\textwidth]{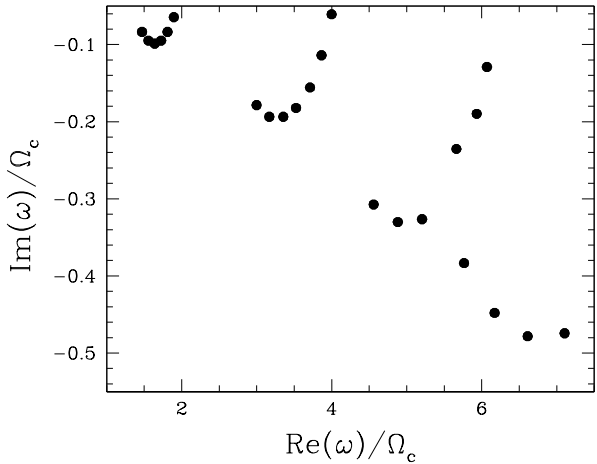}
\caption{Positions of solutions to the dispersion equation for the 
$W=2$ lowered Spitzer plane.  The cluster of modes at 
${\rm Re}(\omega)/\Omega_c\simeq 1.7$ and ${\rm
  Im}(\omega)/\Omega_c\simeq -0.07$ is the same as the modes depicted
in the top panel of Figure \ref{fig:landau}.}
\label{fig:eigen}
\end{figure}

Figure \ref{fig:landau} focuses on modes with frequencies clustered
around ${\rm Re}\left (\omega\right ) = 2\Omega_c$ but with very
different spatial structures.  Thus, they would correspond to the
left-most column in in Figure 4 or the column of modes at
$\omega\simeq 2$ in Figure 2.  In Figure \ref{fig:eigen} we zoom out
for a wider view of the imaginary-$\omega$ half-plane to show
positions of modes clustered about ${\rm Re}\left (\omega\right
)/\Omega_c = 2,\,4,\,6$.

\section{SIMULATIONS}

In this section, we present results from a simple one-dimensional
simulation that illustrates the behaviour predicted by our linear
theory calculations.  Our N-body system comprises
self-gravitating, collisionless infinite sheets.  We use a
particle-mesh scheme in which the density is calculated on a
one-dimensional grid and the force on a sheet at position $z$ is
determined from the integral

\begin{equation}
F(z) = 2\pi G\int dz'\,\rho(z') {\rm sgn}\left (z'-z\right )~.
\end{equation}

\noindent Similar simulations were presented in \citet{weinberg1991, widrow2012}
and \cite{widrow2014}.  

We use initial conditions that correspond to a simple breathing mode
by first setting up an equlibrium distribution and then perturbing the
positions and velocities according to the relations $z = \lambda_z
z_e$ and $v = \lambda_v v_e$ where ($z_e,\,v_e$) are the phase space
coordinates of a particle in the unperturbed system and
$\lambda_{e,v}$ are constants.  With our choice of units, the total
kinetic and potential energies of the unperturbed system are
$T=1/2\pi$ and $V=1/\pi$ respectively.  With this in mind, we choose
$\lambda_v = \left (3 - 2\lambda_z\right )^{1/2}$.  The total energy
of the perturbed and unperturbed systems are therefore the same while
the initial virial ratio for the perturbed system is $R = 2T/V = \left
  (3-2\lambda_z\right )/\lambda_z$.

\begin{figure}
\centering
\includegraphics[width=0.45\textwidth]{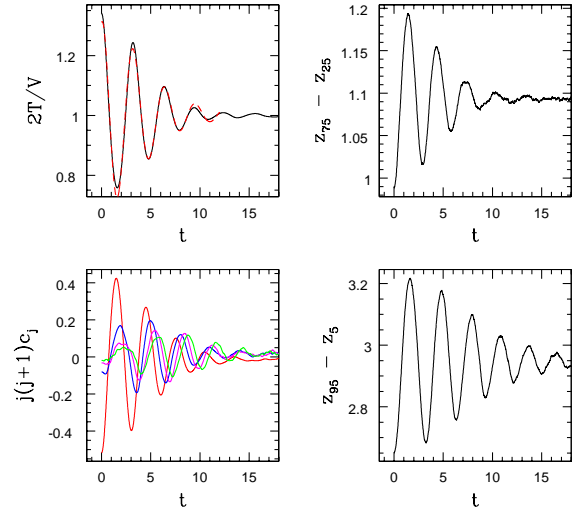}
\caption{Time evolution of a perturbed plane symmetric collisionless
 system.  The upper left panel shows the virial ratio $R\equiv
 2T/V$ as a function of time (black solid curve) as well as a two-term fit based on  
Eq.\,\ref{eq:fitvirial} (red dashed curved).  The lower left panel 
shows the first four even coefficients $j\left
(j+1\right )c_j$ that appear in the expansion for the density. 
Colors are red, blue, magenta, and green for $j=2,\,4,\,6$ and $8$ 
respectively.  The upper right panel shows the time evolution of the width  
of the region that contains the inner $50\%$ of the mass.  The lower 
right panel shows the same for the inner $90\%$ of the mass.}
\label{fig:virial}
\end{figure}

In Figure \ref{fig:virial} we show the time evolution of various
system properties for a simulation with $200K$ particles and
$\lambda_z=0.9$.  For example, in the upper left panel, we show the
virial ratio $R$.  During the initial phase from $t=0$ to $t\simeq
12$ the amplitude of the oscillations in $R$ damps
from an initial value of $0.3$ to $\sim 0.02$.  The time-dependence of
$R$ during this phase can be described by a model comprised of the sum
of damped exponentials:

\begin{equation}\label{eq:fitvirial}
R = R_0 + \sum_{i=1}^n C_i \cos{\left (\omega_i t - \beta_i\right
  )}e^{-\alpha_i t}~.
\end{equation}

\noindent The fit for the run shown in Figure \ref{fig:virial} has two
terms with $\left (\omega_i,\,\alpha_i\right ) \simeq \left
  (1.5,\,0.25\right ),\, \left (2.0,\,0.24\right )$.  Note that we
only fit Eq.\,\ref{eq:fitvirial} for $t<12$.  At later times,
perturbations are driven by the Poisson noise of the simulation (see
below).  The simulation results are in good qualitative agreement with
our analytic results.  In particular, the lower panel of Figure
\ref{fig:landau} shows that the typical modes of the Spitzer sheet
have an oscillation frequency of ${\rm Re}(\omega)\simeq 2$ and an
exponential damping constant of ${\rm Im}(\omega)\simeq 0.2$.
However, because the oscillations damp after only a few cycles
we are unable to resolve the detailed mode structure anticipated in 
Figure \,\ref{fig:landau}.

A key feature of vertical oscilations is that they damp most
rapidly in the inner parts of the distribution.  This is illustrated
in the lower left panel of Figure \ref{fig:virial} where we see that
the damping rate decreases with increasing $j$.  Furthermore, if we
compare the top and bottom panels on the right, we see that the
oscillations of the region containing 50\% of the mass damp more rapidly than
the oscilations of the region containing 90\% of the mass.
\citet{widrow2014} also noticed that the high energy part of the
distribution function was more susceptible to vertical oscillations
that were provoked by a passing satellite.  We can therefore
predict that bulk vertical motions will be strongest among stars
in the high (vertical) energy tail of the phase space distribution.

\begin{figure}
\centering
\includegraphics[width=0.45\textwidth]{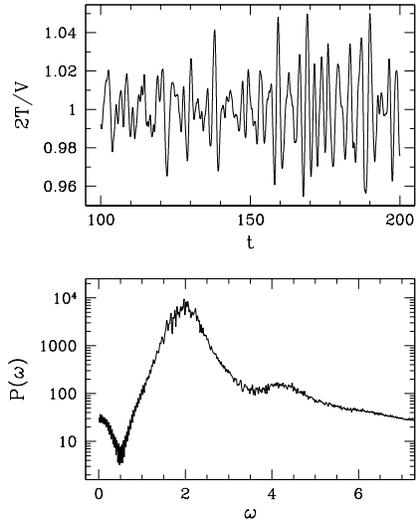}
\caption{Results from a low resolution simulation with 10K particles
  that was run until $t=1300$.  The upper panel shows a small section
of the time evolution of the virial ratio while the lower panel shows
the power spectrum.}
\label{fig:longterm}
\end{figure}

At late times, the initial perturbation has damped away but
oscilations, seeded by Poisson noise due to the finite number of
particles, persist.  In Figure \ref{fig:longterm} we
show late time behaviour of a low-resolution (10K particle) simulation
that was evolved until $t\simeq 1300$.  We show a portion of the time
evolution of the virial ratio (upper panel) and the power spectrum of
the time-domain Fourier transform (lower panel).  We see that there are
continous oscillations with randomly changing phase and amplitude
but with a characteristic frequencies of $\omega\simeq 2$ and
$\omega\simeq 4$.

\section{DISCUSSION AND CONCLUSIONS}

A passing satellite can irrevocably change the properties of a
galactic disk in two fundamental ways.  First, orbital energy from the
satellite can be transferred to disk stars thereby heating and
thickening the disk.  Second, the satellite can provoke the formation
of secular phenomena such as a bar, warp, or spiral structure.  In
either case, the initial response of the disk to the satellite can
involve coherent oscillations.  The subsequent evolution of
these motions is governed by a mix of processes that include the
restoring force of the disk's own self-gravity, Landau damping,
differential rotation, and swing amplification of spiral waves.  In
this paper, we have focused on the first two of these effects by
restricting our attention to plane symmetric systems.  Indeed, the
structure of the oscillatory behaviour for a self-gravitating,
plane-symmetric system is already quite complicated, especially in the
case of a collisionless (i.e., stellar) system.  There, the modes form
a double series defined by two eigenvalues, one that
determines the mode's frequency and the other that determines its
spatial structure

Linear perturbations of both gaseous and stellar plane symmetric
systems divide neatly according to their parity with respect to the
Galactic midplane.  The North-South asymmetries in the number counts
found in \citet{widrow2012} and \citet{yanny2013} by construction
picked out odd parity density modes.  Bending modes, which correspond
local displacements of the disk from the midplane, can also be thought
of as odd parity density perturbations.  The midplane displacements
seen in the interstellar medium \citep{nakanishi2006} are an example
of this.

By contrast modes where the density perturbation has even parity have
bulk velocity fields that are odd in parity.  The simplest example is
the breathing mode perturbation, which may be present in the SEGUE
\citep{widrow2012}, RAVE \citep{williams2013}, and LAMOST
\citep{carlin2013} data sets.  Unfortunately, a clear picture of the
velocity field throughout the 1-2 kiloparsec neighbourhood of the Sun
is lacking in large part because of the complicated and incomplete
footprints of the surveys (see \citet{carlin2013} for a detailed
discussion).  A combined analysis of the three data sets might improve
this situation as will satellite data from Gaia \citep{perryman2001}.
As well, \citet{bovy2014} proposed an alternative way to characterize
the velocity field of the Galactic disk.  The idea is to calculate the
power spectrum of the velocity field after subtracting an axisymmetric
model that accounts for the rotation of the disk.  They find that the
largest contribution to the power spectrum on large scales comes from
the Sun's motion relative to the local standard of rest.  In addition,
they find a broad peak in the power spectrum with $0.2\,{\rm kpc}^{-1}
< k < 0.9\,{\rm kpc}^{-1}$ and the associated motions might be
associated with the time-dependent gravitational potential of the bar.
While their analysis uses only heliocentric line-of-sight velocities
and focuses on the two-dimensional in-plane velocity field, the
technique could easily be extended to include proper motions and
velocities perpendicular to the Galactic plane.

\citet{robin2003} estimate the density of stars and dark matter in the
midplane of the Galaxy at the position of the Sun is $\rho_c \simeq
0.055\,M_\odot \,{\rm pc}^{-3}$.  The period of vertical
oscillations for a star near the midplane is therefore $T_c \simeq
110\,{\rm Myr}$ or approximately one half the period for a star at
this radius to orbit about the center of the Galaxy.  In Section 3, we
found that in the solar neighbourhood, the ratio of the exponential
decay time to the vertical oscillation period for pure vertical modes
typically fell in the range $\tau/T\simeq 2.5-8$.  Thus, these modes
might be expected to persist for $1-4$ orbital periods or
$200-800\,{\rm Myr}$.

Of course, any application of our results to the Milky Way will
require a more detailed understanding of how vertical modes couple to
modes in the disk plane.  As a satellite galaxy passes through the
disk, it excites both bending and breathing modes \citep{widrow2014}.
The former have been implicated in the generation of galactic warps.
\citet{debattista2014} has shown that spiral arms generate compression
and rarefaction in the disk but the reverse seems plausible, namely
that the compression and rarefaction perturbations due to a satellite,
sheared by differential rotation of the disk, generate spiral
structure.  \citet{sellwood2014} have argued that spiral activity in
disk galaxies arise from the superposition of transient unstable
spiral modes.  Satellite galaxies and dark matter subhalos would seem
to provide a natural seed for these instabilities.

\appendix

\section{MATRIX ELEMENTS FOR THE FLUID CASE}

In this appendix, we describe the calculation of the matrix elements
$M_{jk}$ in Eq.\,\ref{eq:matrix}.  As indicated in the text, we
multiply both sides of Eq.\,\ref{eq:Eulerpert} by $-\psi_j$ and
integrate with respect to $z$ from $-\infty$ to $\infty$.  The left
hand side becomes $\omega^2 c_j$.

The first two terms on the right-hand side of Eq.\,\ref{eq:Eulerpert}
can be combined to give

\begin{align}
\int_{-\infty}^\infty  & dz\,\psi_j
\left (\frac{d^2\rho_1}{dz^2} + 
\frac{d\psi_0}{dz}\frac{d\rho_1}{dz}\right )  \\
 & 
= \int_{-1}^1 du\,\left (1-u^2\right )\psi_j\frac{d^2\rho_1}{du^2}
\\
& 
=\sum_k {\cal N}_{jk}
\int_{-1}^1 du\,\left (1-u^2\right )
P_j\frac{d^2}{du^2}
\left (\left (1-u^2\right )P_k\right )
\end{align}
where ${\cal N}_{jk} = k\left (k+1\right )N_j N_k$.  We can then use
Legendre's differential equation as well as the identity
\begin{equation}
\left (1-u^2\right )P'_j(u) = j\left (P_{j-1}(u) - uP_j\right )
\end{equation}
to write this expression in terms of the integrals
\begin{equation}
I_{j_k}^1 = \int_{-1}^1 du\,P_j(u) P_k(u)~,
\end{equation}

\begin{equation}
I_{j_k}^2 \equiv
\int_{-1}^1 du\,u^2P_j(u)P_k(u) 
\end{equation}

\noindent and

\begin{equation}
I_{j_k}^3 \equiv
\int_{-1}^1 du\,uP_j(u)P_{k-1}(u) ~,
\end{equation}
which can be evaluated analytically \citep{arfken2005, weisstein2014}.

The third and fifth terms of Eq.\,\ref{eq:Eulerpert} can be combined
to yield

\begin{align}
\int_{-\infty}^\infty dz\,\psi_j &
\left (
\rho_1\frac{d^2\psi_0}{dz^2} + 
\rho_0\frac{d^2\psi_1}{dz^2}\right )  \\
& = -4\sum_k {\cal N}_{jk} 
\left (I_{j_k}^1 - I_{j_k}^2\right )
\end{align}

\noindent while the fourth term gives

\be
\int_{-\infty}^\infty dz\,\psi_j
\frac{d\rho_0}{dz}\frac{d\psi_1}{dz}{dz}
=  4\sum_k k N_j N_k \left (I^2_{jk} -
  I^3_{jk}\right )~.
\ee

\section{HOMOGENEOUS SLAB}

For the lowest order even parity mode ($j=1$) the potential
can be written as

\be
\psi_1 = \frac{3N_1}{2}\left (E\cos^2\theta - 1\right )
\ee

\noindent while density is constant: $\rho_1 = 3N_1/4\pi$.
Since we are considering an even parity mode, we can write the 
Fourier series for $\psi$ as

\be
\psi_1 = \sum_{n=0}^\infty \psi_{1,n} \cos{n\theta}~.
\ee

\noindent We note that the $n=0$ term does not contribute to the
distribution function.  The relevant Fourier coefficient is then
$\psi_{1,2} = 3N_1 E/4$ distribution function is then 

\be {\rm
  Re}\left (f_1\right ) = \frac{df_0}{dE} \frac{3N_1
  E}{2}\frac{\cos{2\theta}}{2 - \omega^2/4} ~.  
\ee

In order to show that this is indeed a true mode of the system, we
calculate the density directly from the distribution function: $\rho_1
= \int dv\,f_1$.  As discussed in the text, this integral formally
diverges but can be handled by considering a family of equilibrium
distribution functions $f_\lambda(E)$ that approximate $f_0$ but are
continuous at $E=1$.  We then have

\be 
\int dv\,v^2 \frac{df_\lambda}{dE} = \int
dv\,v\frac{df_\lambda}{dv} = -\int dv\,f_\lambda = -\rho_0 
\ee 

\noindent where at the last equality, we let $\lambda\to \infty$.
Moreover, the term proportional to $z^2$ vanishes.  To see this, write
$df_0/dE = \left (1/2z\right )df_0/dz$ and perform the integration
over velocities before differentiating.  The net result is that

\be
\rho_1 = \frac{3N_1}{8\pi}\frac{1}{2 - \omega^2/4}~,
\ee

\noindent which gives $\omega = 6^{1/2}$
\citep{antonov1971, kalnajs1973, fridman1984}.

The next mode ($j=3$) is

\be
\psi_3 = N_3\left (P_4(z) - P_2(z)\right ) = 
\frac{7N_3}{8}\left (5z^2 - 1\right )\left (z^2 - 1\right ) ~.
\ee

\noindent Writing out in terms of trig functions

\be\label{eq:rho3}
\psi_3  = \frac{7N_3}{8}
\left (5E^2\cos^4\theta - 6E\cos^2\theta
+ 1\right )~.
\ee

\noindent With the help of various trigonometric identities, we find

\be 
\psi_{3,4} = \frac{7N_3}{8}\frac{5E^2}{8} 
\ee

\noindent and 

\be
\psi_{3,2} = \frac{7N_3}{8}\left (\frac{5E^2}{2}-3E\right )~.
\ee

\indent Thus, for the distribution function, we have

\be
Re\left (f_3\right ) = \frac{df_0}{dE}\left ( 
\frac{2\psi_{3,2}}{2 - \omega^2/4}\cos{2\theta}+
\frac{2\psi_{3,4}}{2 - \omega^2/16}
\cos{4\theta}\right )~.
\ee

\noindent A straightforward, but tedious calculation similar to the
one performed above leads to an expression for $\rho_3$, which, when
combined with Eq.\,\ref{eq:rho3}, yields a quadratic equation for
$\omega^2$ whose solutions are $\omega/\Omega_c = 2.07,\,3.83$.

\section{DYNAMICS IN THE COMPLEX ENERGY PLANE}

For the Spitzer sheet, we can find the density as a function of the potential 
by integrating the distribution function (Eq.\,\ref{eq:spitzerdf}) over $v$:

\begin{equation}
\rho\left (\psi\right ) = \int_{-\infty}^\infty dv\,f(E)
= \frac{1}{2\pi} e^{-\psi}~.
\end{equation}
 
\noindent The Poisson equation $d^2\psi/dz^2 = 2\exp{\left
    (-\psi\right )}$ can then be integrated to obtain as expression
for the force as a function of the potential:

\begin{equation}
F\left (\psi\right ) = -\frac{d\psi}{dz} = -2\left (1-e^{-\psi}\right )^{1/2}~.
\end{equation}

\noindent The period for a particle with energy $E$ is
\begin{align}\label{eq:complexperiod}
T(E) & = -4\int_{v_{\rm max}}^{v(t)} \frac{dv'}{F\left (\psi\right
 )}\\
& = 2^{3/2}E^{1/2}
\int_0^{\pi/2} 
\frac{\cos{\varphi}d\varphi}
{\left (1-e^{-E\cos^2\varphi}\right )^{1/2}}
\end{align}

\noindent where $v_{\rm max}$ is the maximum velocity of the particle
along its orbit \citep{araki1985}.  In deriving the second expression,
we write $v$ and $\psi$ in terms of the parameter $\varphi$:

\begin{equation}
v = \left (2E\right )^{1/2} \sin{\varphi}~~~~~~~
\psi = E\cos^2{\varphi}
\end{equation}

More generally, we can write $t$ along the first quarter of the
particle's orbit as a function of $\varphi$:

\begin{equation}\label{eq:complextofpsi}  
t\left (\varphi,\,E\right )  = \left (\frac{E}{2}\right )^{1/2}
\int_0^{\varphi(t)}
\frac{\cos\varphi' d\varphi' }
{\left (1-e^{-E\cos^2\varphi'}\right )^{1/2}}
\end{equation}
 
Similarly

\begin{equation}\label{eq:complexzofpsi}
z\left (\varphi,\,E\right )  = E
\int_0^{\varphi(t)}
\frac{\sin{\varphi'}\cos{\varphi'} d\varphi' }
{\left (1-e^{-E\cos^2{\varphi}'}\right )^{1/2}}
\end{equation}

For the lowered Spitzer sheet, $\rho(\psi)$ is given by

\begin{equation}
\rho(\psi) = f_We^{-\psi}\left (
\left (2\pi\right )^{1/2} {\rm erf}\left(  y\right ) - 
2^{3/2} ye^{-y^2}\right )
\end{equation}
\noindent where $y = \left (W-\psi\right )^{1/2}/\sigma$.
As before, we integrate the Poisson equation and find

\begin{equation}
F(\psi) = \left (8\pi\right )^{1/2}
\left ( {\cal F}\left (y_0 \right ) - {\cal F}\left
    (y \right ) \right )^{1/2}
\end{equation}

\noindent where $y_0 \equiv W^{1/2}$ and 

\begin{equation}
{\cal F}(z) = \left (2\pi \right )^{1/2} e^{z^2}{\rm erf} z 
- 2^{3/2} \left (z + \frac{2z^3}{3}\right )
\end{equation}

Finally, we write the complex Fourier coefficients of 
the basis functions $\psi_j$ as

\begin{equation}
\psi_{nj}(E) = 
\frac{1}{T(E)}
\int_0^{2\pi} d\varphi\,\frac{dt}{d\varphi} \,
\psi_j\left (z\left (E,\,\theta\right ) \right ) e^{-in\theta}
\end{equation}

\noindent where $z$, $t$, and $\theta$ are complex
functions of $E$ and of the real integration variable $\varphi$
through Eqs.\,\ref{eq:complextofpsi} and \ref{eq:complexzofpsi}.

\label{lastpage}

\end{document}